\documentclass[a4paper]{article}
\usepackage{xcolor}
\usepackage[utf8]{inputenc}
\usepackage[T1]{polski}
\usepackage{graphicx}\graphicspath{{img/}}
\usepackage{lmodern,fancyhdr,secdot,float,amsmath,amsfonts,amsthm,amssymb}
\usepackage{booktabs,tabularx}
\usepackage[margin=2cm]{geometry}
\usepackage{multicol}
\usepackage[hidelinks,unicode]{hyperref}
\usepackage{titlesec}
\titleformat{\section}{\centering\normalsize\bf}{\thesection.}{.5em}{\MakeUppercase}
\titleformat*{\subsection}{\bf\normalsize\selectfont}
\titleformat*{\subsubsection}{\bf\normalsize\selectfont}
\newcommand{\titlePL}[1]{\large\textbf{ #1}}
\newcommand{\titleEN}[1]{\normalsize #1}
\newcommand{\keywordsPL}[1]{\small\textbf{Słowa kluczowe:} #1}
\newcommand{\keywordsEN}[1]{\small\textbf{Keywords:} #1}
\newcommand{\abstractPL}[1]{\small\textbf{Streszczenie:} #1}
\newcommand{\abstractEN}[1]{\small\textbf{Abstract:} #1}

\definecolor{logo_color}{RGB}{40, 69, 166}

\begin{document}\thispagestyle{empty}\pagestyle{fancy}
\begin{minipage}[t]{0.5\textwidth}\vspace{0pt}%
\includegraphics[scale=0.9]{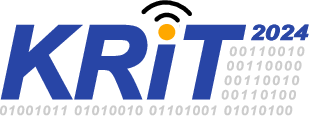}
\end{minipage}
\begin{minipage}[t]{0.45\textwidth}\vspace{12pt}%
\centering
\color{logo_color} KONFERENCJA RADIOKOMUNIKACJI\\ I TELEINFORMATYKI\\ KRiT 2024
\end{minipage}

\vspace{1cm}

\begin{center}
\titlePL{FORMOWANIE KLASTRÓW OBSŁUGUJĄCYCH W SIECI 6G OPEN RAN O ARCHITEKTURZE ZORIENTOWANEJ NA UŻYTKOWNIKA}

\titleEN{SERVING CLUSTER FORMULATION IN OPEN RAN USER-CENTRIC CELL-FREE NETWORK}\medskip

Marcin Hoffmann $^{1}$;

\medskip

\begin{minipage}[t]{0.6\textwidth}
\small $^{1}$ Politechnika Poznańska, Poznań, \href{mailto:email}{marcin.hoffmann@put.poznan.pl}\\
\end{minipage}

\end{center}

\medskip

\begin{multicols}{2}
\noindent
\abstractPL{
Jednym z najważniejszych wyzwań związanych z implementacją sieci Open RAN o architekturze zorientowanej na użytkownika jest odpowiednie formowanie tzw. klastrów obsługujących. Problem ten może być efektywnie rozwiązany za pomocą zaproponowanego algorytmu wykorzystującego współpracę pomiędzy xApp'em i rApp'em. Wyniki symulacji pokazały, że zaproponowane rozwiązanie pozwala na zwiększenie mediany z rozkładu przepływności użytkowników o 16\% względem standardowej architektury zorientowanej na sieć.\footnote[1]{Praca powstała w~ramach projektu PRELUDIUM finansowanego przez Narodowe Centrum Nauki nr 2022/45/N/ST7/01930.}}
\medskip

\noindent
\abstractEN{
One of the main challenges associated with the implementation of an Open RAN User-Centric Cell-Free network is the appropriate formulation of serving clusters. This problem can be effectively solved using a proposed algorithm that leverages cooperation between xApp and rApp. Simulation results have shown that the proposed solution allows for a 16\% increase in the median of the user throughput distribution compared to the state-of-the-art network-centric architecture.}
\medskip

\noindent
\keywordsPL{6G, Massive MIMO, sieć zorientowana na użytkownika, Open RAN}
\medskip

\noindent
\keywordsEN{6G, Massive MIMO, User-Centric networks, Open RAN}

\section{Wstęp}

Jedną z kluczowych cech sieci mobilnych szóstej generacji (6G) jest zastosowanie tzw. architektury zorientowanej na użytkownika (\textit{ang. User-Centric Cell-Free}) w połączeniu\newline z techniką M-MIMO (\textit{ang. Massive Multiple-Input Multiple-Output})~\cite{yang2021}. Poprzednie generacje sieci mobilnych stosowały tzw. architekturę zorientowaną na sieć, gdzie użytkownik typowo obsługiwany jest przez jedną stację bazową od której odbiera najwyższą moc sygnału referencyjnego RSRP (\textit{ang. Reference Signal Received Power}). W nowej architekturze zorientowanej na użytkownika może być on obsługiwany przez wiele zsynchronizowanych stacji bazowych (nazywanych w tym kontekście punktami dostępu) koordynowanych przez tzw. centralną jednostkę przetwarzania CPU (\textit{ang. Central Processing Unit}). Wykorzystanie takiej architektury pozwala przede wszystkim na zmniejszenie dysproporcji w mocy odbieranej w różnych punktach sieci. Jest to szczególnie istotne w przypadku użytkowników znajdujących się w podobnej odległości od kilku stacji bazowych~\cite{ngo2015}. W tzw. kanonicznej implementacji sieci zorientowanej na użytkownika zakłada się, że wszystkie stacje bazowe będą obsługiwały wszystkich użytkowników~\cite{buzzi2017}. W praktyce jednak, zarówno stacje bazowe jak i CPU mają ograniczone zasoby obliczeniowe i mogą obsłużyć jedynie skończoną liczbę użytkowników, np. ze względu na konieczność estymacji kanału i obliczania wag prekodera. Ponadto, część stacji bazowych znacząco oddalonych od użytkownika będzie miała marginalny wpływ na poprawę jego przepływności, "marnując" w ten sposób cenne zasoby radiowe. Z tej perspektywy jednym z kluczowych zagadnień dla sieci zorientowanej na użytkownika jest procedura tzw. formowania klastrów obsługujących. Ma ona na celu określenie zbioru stacji bazowych, które powinny obsługiwać konkretnego użytkownika~\cite{ammar2022}. Klastry obsługujące mogą być dynamicznie formowane np. na bazie RSRP, z wykorzystaniem różnych algorytmów np. uczenia maszynowego ML (\textit{ang. Machine Learning}). Ponadto, formowanie klastrów obsługujących może uwzględniać różne funkcje celu, takie jak: maksymalizacja efektywności energetycznej lub maksymalizacja przepływności osiąganych przez użytkowników.

Potencjalną szansą na praktyczną implementację sieci zorientowanej na użytkownika jest wykorzystanie koncepcji otwartej sieci dostępowej Open RAN (\textit{ang. Open Radio Access Network})~\cite{Ranjbar2022}. W ramach koncepcji Open RAN zaproponowany został tzw. podział 7.2, pozwalający na rozdzielenie funkcji stacji bazowej na zwirtualizowaną jednostkę rozproszoną O-DU (\textit{ang. Open RAN Distributed Unit}) realizującą m.in., funkcje warstwy fizycznej wyższego poziomu i warstwy dostępu do łącza, oraz fizyczną jednostkę radiową O-RU (\textit{Open RAN Radio Unit}), odpowiedzialną za funkcje warstwy fizycznej niższego poziomu i bezpośrednio transmisje radiową. Korzystając z podziału 7.2, tzw. węzeł E2 (tutaj O-DU) może służyć jako CPU i kontrolować transmisję z wielu jednostek radiowych O-RU wykorzystujących technikę M-MIMO. Jedną z kluczowych cech sieci Open RAN jest możliwość kontrolowania sieci dostępowej poprzez implementację dedykowanych algorytmów w formie tzw. xApp'ów i rApp'ów, które są wdrażane w ramach tzw. inteligentnego kontrolera sieci dostępowej pracującego w czasie bliskim rzeczywistemu Near-RT RIC (\textit{ang. Near Real-Time RAN Intelligent Controller}), oraz tzw. inteligentnego kontrolera sieci dostępowej pracującego w czasie nierzeczywistym Non-RT RIC. Near-RT RIC pracuje w pętli kontroli o skali czasu od 10~ms do 1~s, podczas gdy Non-RT RIC w pętli kontroli o skali czasu powyżej 1~s. Wykorzystanie xApp'u do formowania klastrów obsługujących w sieci Open RAN zorientowanej na użytkownika zostało zaproponowane w~\cite{Beerten2022}. Autorzy jednak rozważali jedynie wykorzystanie Near-RT RIC'a, bez uwzględnienia dodatkowych danych, które mogłyby być przesyłane przez rApp w Non-RT RIC'u. Ponadto, analiza symulacyjna przeprowadzona w tej pracy, zakłada prosty model kanału radiowego, oraz wąskopasmową transmisję na pojedynczej nośnej, podczas gdy rzeczywiste sieci 5G, a prawdopodobnie także 6G, wykorzystują wielodostęp oparty na modulacji wielotonowej OFDMA (\textit{ang. Orthogonal Frequency Division Multiple Access}).

W tej pracy proponowany jest algorytm formowania klastrów obsługujących w sieci Open RAN o architekturze zorientowanej na użytkownika z wykorzystaniem dwóch współpracujących ze sobą aplikacji: xApp'a w Near-RT RIC'u i rApp'a w Non-RT RIC'u. Zaproponowane rozwiązanie zostanie zbadane za pomocą zaawansowanego symulatora sieci M-MIMO o architekturze zorientowanej na użytkownika, wykorzystującego trójwymiarowe śledzenie promieni (\textit{ang. 3D Ray Tracing}), zaproponowanego we wcześniejszej pracy autora~\cite{hoffmann2024}. W dalszych częściach tej pracy, w~Rozdziale~\ref{sec:system_model} przedstawiona zostanie rozważana sieć Open RAN o architekturze zorientowanej na użytkownika. Rozdział~\ref{sec:xapprapp} zawiera opis zaproponowanego algorytmu formowania klastrów obsługujących. Wyniki symulacji przedstawione są w~Rozdziale \ref{sec:results}, a~wnioski zostały sformułowane w~Rozdziale~\ref{sec:conclusions}.

\section{Sieć Open RAN o Architekturze Zorientowanej na Użytkownika} \label{sec:system_model}

\begin{figure}[H]
\centering
\includegraphics[width=0.49\textwidth]{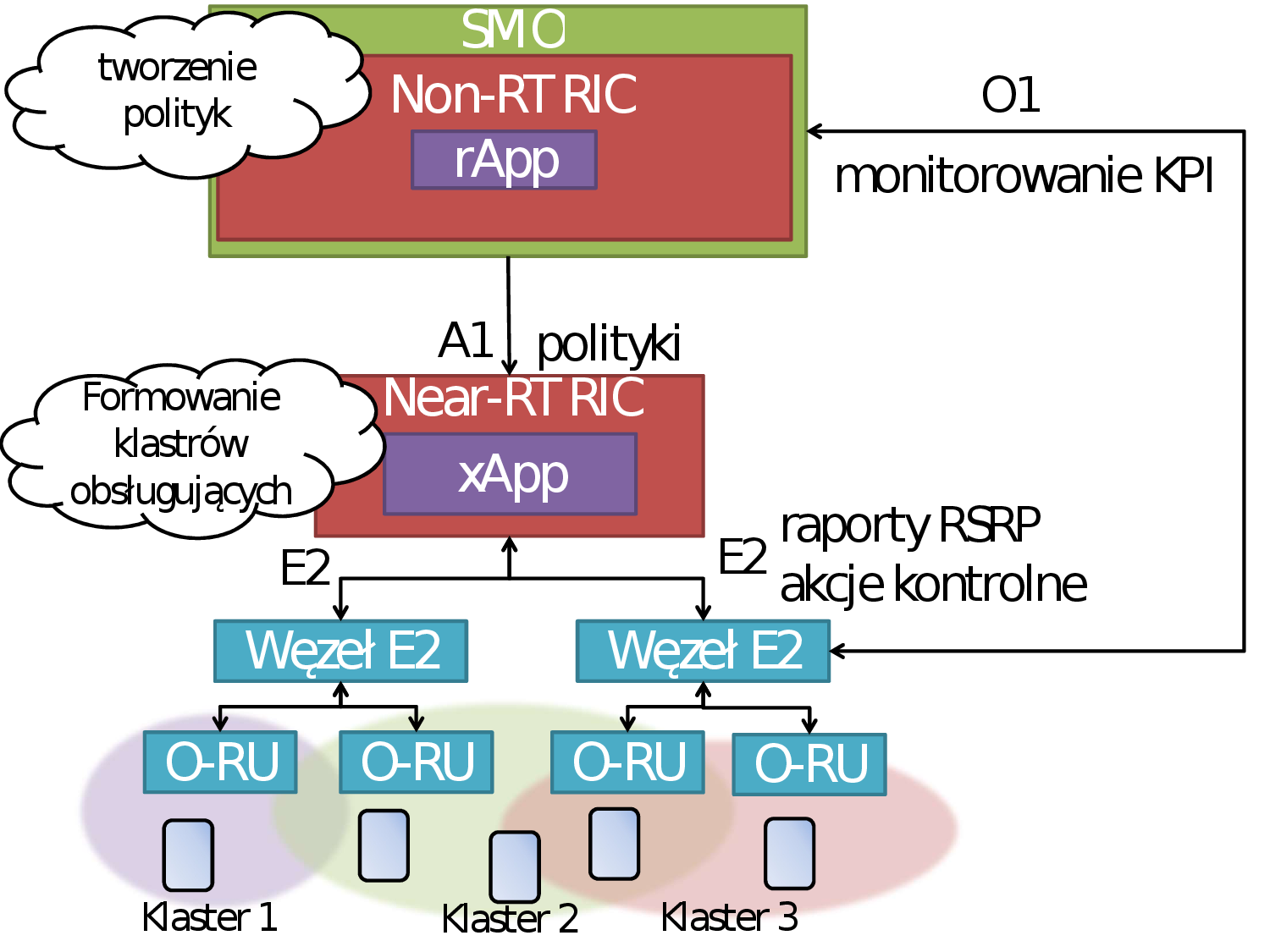}
\caption{Sieć Open RAN o architekturze zorientowanej na użytkownika.}
\label{fig:system_model}
\end{figure}
W tej pracy rozważane jest łącze w dół w sieci Open RAN zorientowanej na użytkownika przedstawionej na Rys.~\ref{fig:system_model}. W tej sieci O-RU jest jednostką radiową wyposażoną w macierz antenową M-MIMO. O-RU odpowiada bezpośrednio z transmisję danych użytkownika w interfejsie radiowym. Węzeł E2 to połączona zwirtualizowana jednostka rozproszona O-DU i zwirtualizowana jednostka centralna O-CU (\textit{ang. Open RAN Centralized Unit}). Pełni ona rolę CPU, tzn., odpowiada za przydział zasobów radiowych, a także koordynację transmisji do użytkowników z kontrolowanych przez nią fizycznych jednostek radiowych O-RU. Jednym z najistotniejszych zadań węzła E2 w ramach koordynacji transmisji, jest zestawienie klastra obsługującego dla każdego z użytkowników. Wymaga ono konfiguracji jednostek radiowych O-RU tworzących ten klaster w celu zapewnienia, m.in., wymiany danych użytkownika, synchronizacji transmisji, oraz wymiany informacji o współczynnikach kanałów radiowych poszczególnych użytkowników. W szczególności, potencjalnym wyzwaniem jest formowanie klastrów obsługujących wymagających koordynacji pomiędzy dwoma węzłami E2, tj. fragmentami sieci obsługiwanymi przez różne CPU. Formowanie klastrów obsługujących może być kontrolowane z użyciem interfejsu E2 przez xApp znajdujący się w Near-RT RIC'u. W rozważanej sieci taka kontrola polega na wskazaniu węzłowi E2 które jednostki radiowe O-RU powinny obsługiwać danego użytkownika. Oprócz przesyłania danych kontrolnych, interfejs E2 umożliwia również przesyłanie raportów dotyczących tzw. kluczowych wskaźników wydajności KPI (\textit{ang. Key Performance Indicators}) sieci dostępowej i użytkowników, np. wartości RSRP, które xApp może wykorzystać w procesie formowania klastrów obsługujących. Z drugiej strony węzeł E2 połączony jest z tzw. SMO (\textit{ang. Service Management and Orchestration}) interfejsem O1, który umożliwia m.in., przekazywanie raportów zawierających KPI, np. rozkład przepływności użytkowników. Tego typu dane mogą być używane przez rApp do formowania tzw. polityk (\textit{ang. policies}) mających na celu kierowanie xApp'em i docelowo poprawę wydajności sieci dostępowej. Polityki przekazywane są poprzez interfejs A1 z Non-RT RIC'a do Near-RT RIC'a. W rozważanej sieci polityka może zawierać parametry algorytmu formowania klastrów obsługujących ustalone na podstawie analizy obserwowanych KPI w dłuższej skali czasowej.

\section{Algorytm Formowania Klastrów Obsługujących} \label{sec:xapprapp}

W tej pracy proponowany jest algorytm formowania klastrów obsługujących w sieci Open RAN zorientowanej na użytkownika z wykorzystaniem współpracy pomiędzy xApp'em i rApp'em. W dalszej części tego rozdziału opisana zostanie najpierw rola xApp'a, a następnie rola rApp'a w zaproponowanym algorytmie.

\subsection{Rola xApp'a}

Rolą xAppa w zaproponowanym podejściu jest dynamiczne formowanie klastrów obsługujących w oparciu o wartości RSRP poszczególnych użytkowników raportowane poprzez interfejs E2, oraz parametry algorytmu otrzymywane z Non-RT RIC'a za pośrednictwem interfejsu A1. W pierwszym kroku xApp wyznacza zbiór jednostek radiowych O-RU, które powinny obsługiwać danego użytkownika. Dla $k$-tego użytkownika, niezależnie spośród $L$ jednostek radiowych O-RU indeksowanych przez zmienną~$l$, wybieranych jest $N_k$ o największej wartości RSRP, które zapewnią spełnienie warunku \cite{tugfe2021cellfree}:
\begin{equation}\label{eq:serving_cluster_formulation}
\frac{\sum_{l \in \mathcal{M}_k}p_{k,l}}{\sum_{l=1}^{\mathrm{L}}p_{k,l}} \geq \Delta,
\end{equation}
gdzie $\mathcal{M}_k$ jest klastrem obsługującym $k$-tego użytkownika, $p_{k,l}$ oznacza RSRP odebrane przez $k$-tego użytkownika od $l$-tej jednostki radiowej O-RU, a $\Delta$ jest parametrem algorytmu przekazywanym przez rApp. Po wyznaczeniu zbioru jednostek radiowych O-RU obsługujących $k$-tego użytkownika - $\mathcal{M}_k$, jest on przesyłany przez xApp jako akcja kontrolna do węzła E2. 

\subsection{Rola rApp'a}

W zaproponowanym algorytmie, rolą rApp'a znajdującego się w Non-RT RIC'u jest tworzenie polityk, które kierują procesem formowania klastrów obsługujących w xApp'ie. Polityka wysyłana jest do Non-RT RIC'a za pomocą interfejsu A1 i zawiera wartość parametru~$\Delta$, która powinna być zastosowana przez xApp. W tej pracy zaproponowano podejście, w którym wybór parametru $\Delta$ dokonywany jest z użyciem ML i zdefiniowany jako tzw. problem wielorękiego bandyty (\textit{ang. Multi-Arm Bandit})~\cite{Sutton1998}. W tym przypadku każda "ręka" wielorękiego bandyty odpowiada jednej wartości parametru $\Delta$. W równych odstępach czasowych $T$ rApp podejmuje akcję - wybiera jedną z wartości parametru $\Delta$ i przekazuje ją do xApp'a za pośrednictwem interfejsu A1. Zaproponowane jest, żeby algorytm podejmował akcje w sposób zachłanny tzn. zawsze wybierał wartość parametru $\Delta$ związaną z najwyższą spodziewaną nagrodą, oznaczoną jako $Q(\Delta)$. Następnie, obserwuje związaną z nią nagrodę - proponujemy, żeby była nią mediana z rozkładu przepływności użytkowników otrzymanego w okresie $T$, następującym po wyborze akcji. W tym celu rApp monitoruje odpowiednie KPI za pośrednictwem interfejsu O1. Po obliczeniu aktualnej nagrody rApp aktualizuje wartość spodziewanej nagrody $Q(\Delta)$ według wzoru:
\begin{equation}
    Q(\Delta) \leftarrow \alpha r + (1-\alpha)Q(\Delta),
\end{equation}
gdzie $r$ oznacza zaobserwowaną nagrodę, a $\alpha$ jest arbitralnie dobieranym parametrem w zakresie $<0:1>$. Po aktualizacji nagrody rApp wybiera kolejną akcję i cykl się powtarza.

\section{Wyniki Symulacji} \label{sec:results}
\begin{figure}[H]
\centering
\includegraphics[trim={0cm 0cm 4cm 1cm},clip,width=0.49\textwidth]{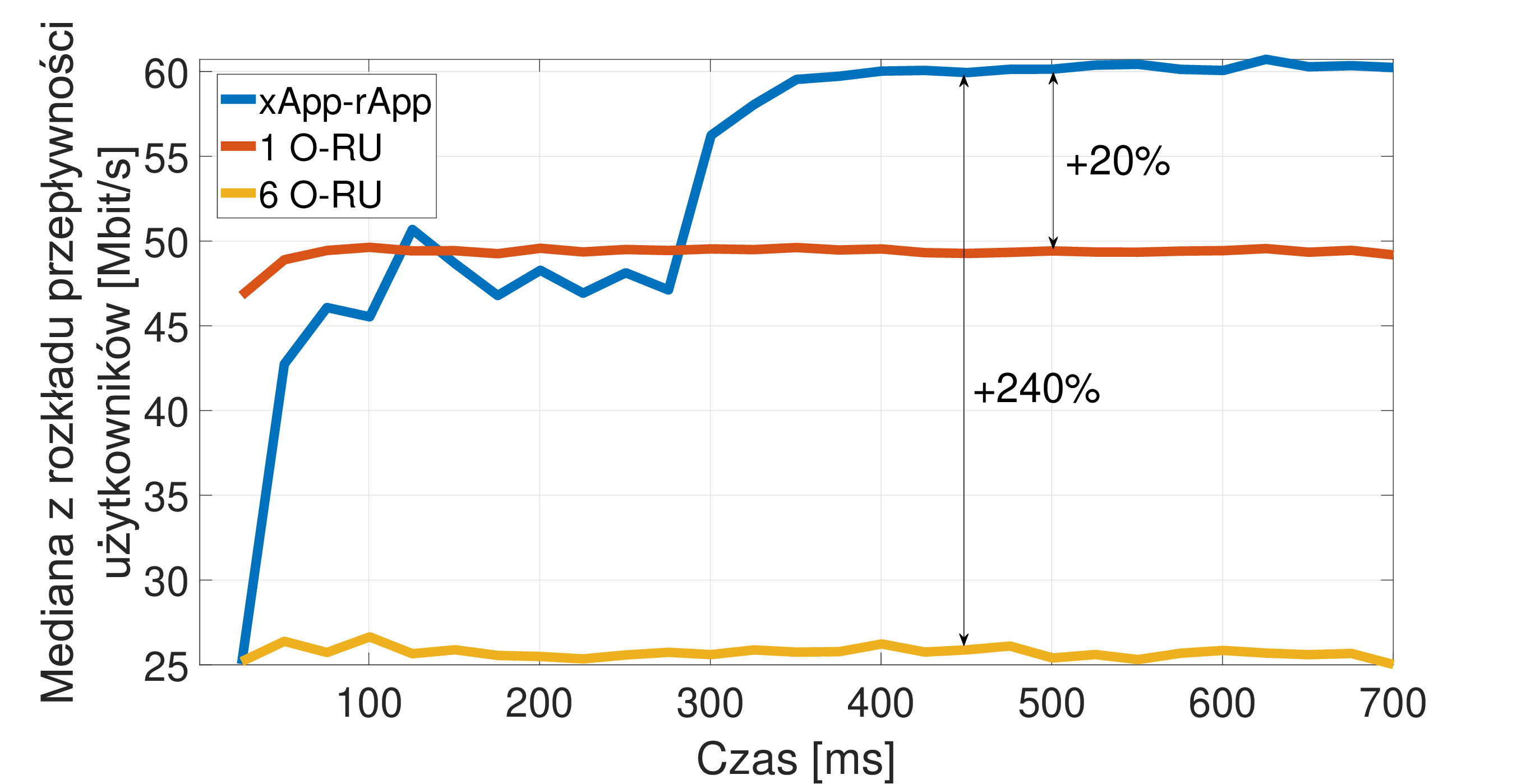}
\caption{Mediana z rozkładu przepływności użytkowników w funkcji czasu, uśredniona po 15 przebiegach symulacyjnych.}
\label{fig:learning}
\end{figure}

\begin{figure}[H]
\centering
\includegraphics[trim={2cm 0cm 1cm 1cm},clip,width=0.49\textwidth]{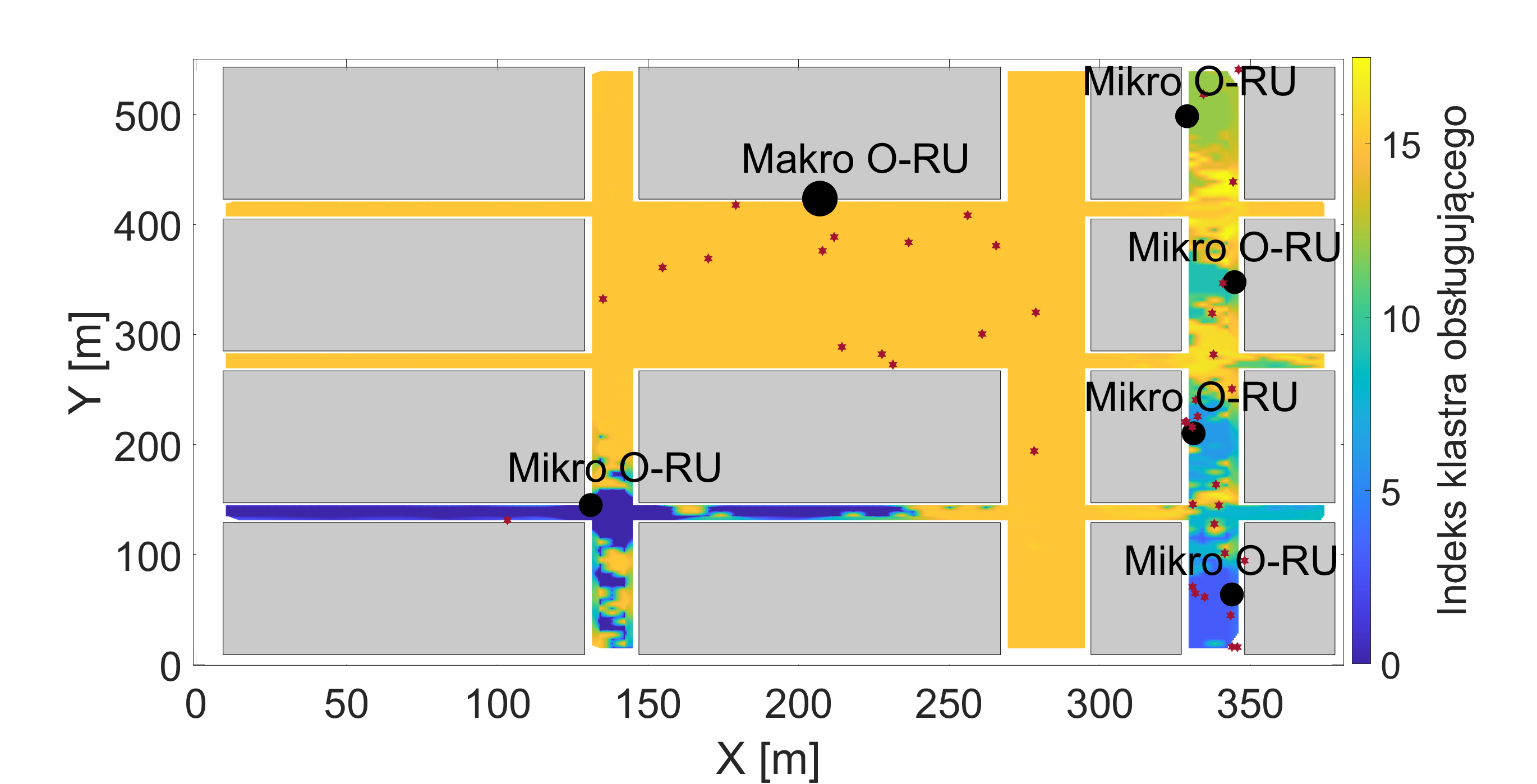}
\caption{Rozmieszczenie jednostek radiowych O-RU, przykładowe rozmieszczenie użytkowników (ciemnoczerwone gwiazdy), oraz podział sieci na klastry obsługujące dla $\Delta=0.8$.}
\label{fig:clusters}
\end{figure}

Zaproponowany algorytm formowania klastrów obsługujących w sieci Open RAN zorientowanej na użytkownika został zbadany za pomocą zaawansowanego symulatora zaproponowanego przez autora w poprzedniej pracy~\cite{hoffmann2024}. Symulator modeluje sieć M-MIMO o architekturze zorientowanej na użytkownika i wykorzystującej OFDMA. Zastosowano w nim realistyczny model kanału radiowego oparty o trójwymiarowe śledzenie promieni firmy Wireless InSite, jak również podział na wiele bloków funkcjonalnych takich jak: formowanie klastrów obsługujących, dedykowany algorytm przydziału zasobów radiowych, prekoder ZF (\textit{ang. Zero-Forcing}) działający niezależnie dla każdej jednostki radiowej O-RU i 15 schematów modulacji i kodowania. W ramach jednej jednostki radiowej O-RU, podział mocy w obrębie pojedynczego bloku zasobów odbywa się równo pomiędzy wszystkich użytkowników, którym został on przydzielony w rezultacie multipleksacji przestrzennej. Pozostałe parametry przeprowadzonych symulacji zostały zestawione w Tabeli~\ref{tab1}. Przeprowadzono 15 przebiegów symulacyjnych, trwających 700~ms (1400 slotów czasowych), w których decyzja o zmianie parametru $\Delta$ podejmowana była przez w odstępach czasowych $T=25$~ms. Każdy przebieg symulacyjny związany był z niezależnym wylosowaniem 40 użytkowników. Dodatkowo zaproponowane rozwiązanie porównano z architekturą zorientowaną na sieć (1 O-RU), gdzie każdy użytkownik obsługiwany jest przez jedna jednostkę radiową O-RU, oraz z tzw. kanoniczną implementacja architektury zorientowanej na użytkownika (6 O-RU), gdzie wszystkie jednostki radiowe O-RU obsługują wszystkich użytkowników. Na Rys.~\ref{fig:learning} przedstawiono medianę z rozkładu przepływności użytkowników uśrednioną po 15 przebiegach symulacyjnych w funkcji czasu. Widać, że w przypadku zaproponowanego algorytmu bazującego na xApp'ie i rApp'ie, przez około 300~ms (12 cykli decyzyjnych) rApp uczy się odpowiedniego doboru parametru $\Delta$, by później ustabilizować osiąganą medianę przepływności na poziomie około 60 Mbit/s, zapewniając $20$\% zysku względem architektury zorientowanej na sieć i 240\% względem kanonicznej implementacji architektury zorientowanej na użytkownika. Wyniki te wskazują na konieczność odpowiedniego formowania klastrów obsługujących, gdyż w przypadku gdy wszystkie jednostki radiowe obsługują wszystkich użytkowników, często duża część mocy nadawana jest do bardzo odległych użytkowników. 
\begin{table}[H]
\centering
\caption{Parametry Symulacji}
\label{tab1}
\medskip
\begin{tabularx}{1.0\columnwidth}{p{0.65\linewidth} p{0.17\linewidth} }
\toprule
\textit{Parametr} & \textit{Wartość}\\ \midrule
Częstotliwość nośna		& 3.6~GHz \\ \hline
Liczba bloków zasobów & 69 \\ \hline
Liczba węzłów E2 & 1 \\ \hline
Liczba jednostek radiowych O-RU & 1 makro, 5 mikro \\ \hline
Wysokość anten - makro O-RU & 45~m\\ \hline
Wysokość anten - mikro O-RU & 6~m\\ \hline
Moc nadawana - makro O-RU & $46$~dBm \\ \hline
Moc nadawana - mikro O-RU & $30$~dBm \\ \hline
Liczba anten - makro O-RU & 128\\ \hline
Liczba anten - mikro O-RU & 32 \\ \hline
Czas symulacji & 700~ms \\ \hline
Liczba użytkowników & 40 \\ \hline
Wireless InSite - liczba odbić & 15 \\ \hline
Wireless InSite - liczba dyfrakcji & 1\\\bottomrule
\end{tabularx}
\end{table}
Na Rys~\ref{fig:clusters} przedstawione zostało rozmieszczenie jednostek radiowych O-RU, przykładowe rozmieszczenie użytkowników, oraz podział rozważanej sieci na klastry obsługujące dla wartości parametru $\Delta=0.8$, która była najczęściej wybierana przez rApp podczas przeprowadzonych symulacji. Kolor odpowiada zbiorowi jednostek radiowych O-RU obsługujących użytkownika znajdującego się w danym punkcie. Duża część rozważanej sieci jest pokrywana przez najsilniejszą jednostkę radiową makro O-RU, stąd największy obszar zajmuje klaster pomarańczowy, w którym użytkownicy obsługiwani są wyłącznie przez tę jednostkę. Dzięki temu jednostki radiowe mikro O-RU nie zużywają mocy na obsługę użytkowników znacząco od nich oddalonych lub oddzielonych budynkami. Można natomiast zaobserwować, że w sąsiedztwie jednostek radiowych mikro O-RU pojawia się wiele klastrów obsługi. Jest to szczególnie widoczne w alejce po prawej stronie. Sformowane klastry pozwalają na obsługę użytkowników przez sąsiadujące ze sobą jednostki radiowe O-RU i tym samym zwiększenie mocy odbieranej przez znajdujących się tam użytkowników, oraz koordynacje interferencji.
\begin{figure}[H]
\centering
\includegraphics[trim={2cm 0.0cm 4cm 1.3cm},clip,width=0.49\textwidth]{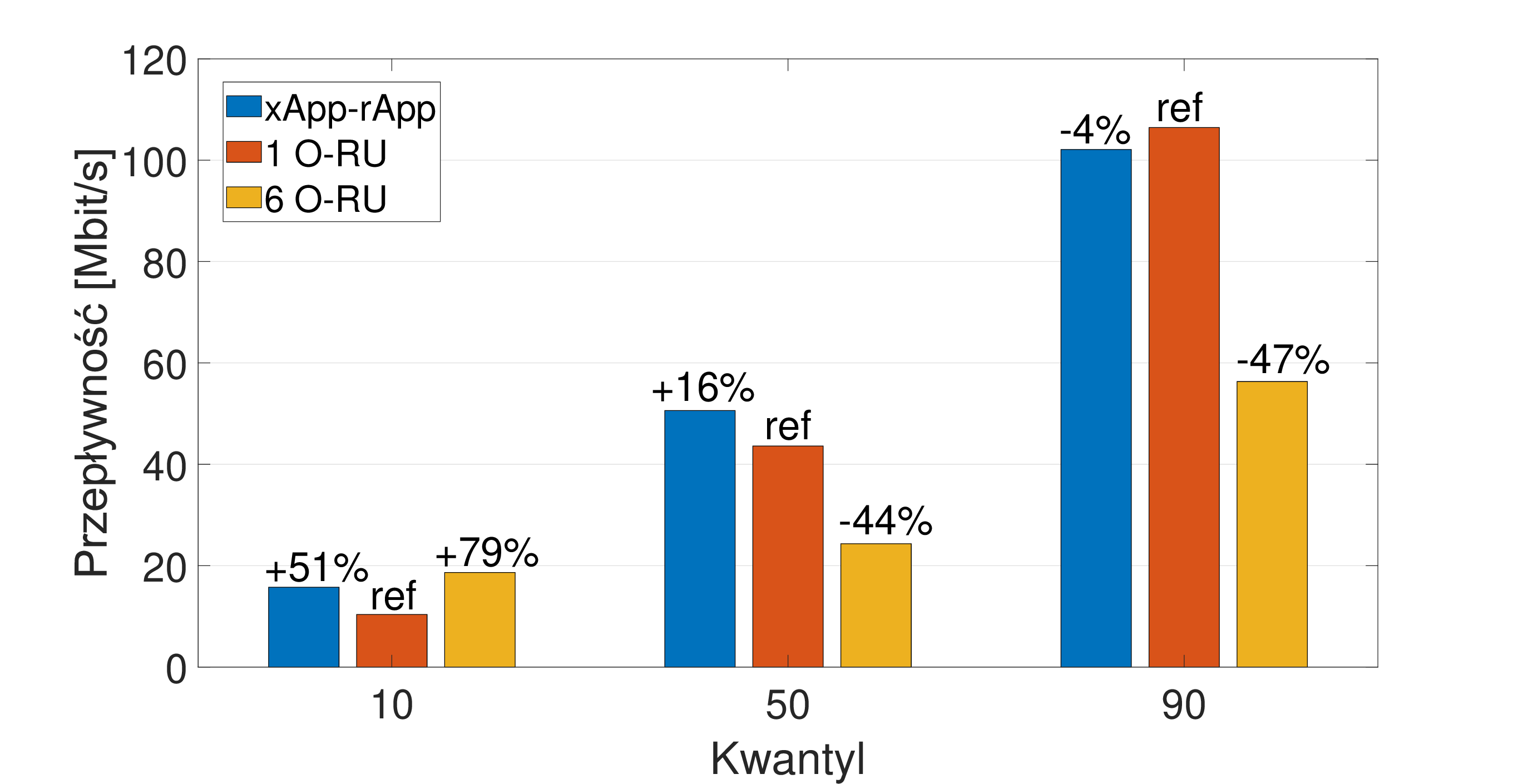}
\caption{Kwantyle rzędu 10, 50 i 90 z rozkładu przepływności użytkowników.}
\label{fig:bar_rate}
\end{figure}

Na Rys.~\ref{fig:bar_rate} przedstawione zostały kwantyle rzędu 10, 50 i 90 z rozkładu przepływności użytkowników otrzymanego z łącznie 15 przebiegów symulacyjnych, dla zaproponowanego algorytmu formowania klastrów obsługujących bazującego na współpracy pomiędzy xApp'em i rApp'em, architektury zorientowanej na sieć i kanonicznej implementacji architektury zorientowanej na użytkownika. Widać, że dla kanonicznej implementacji architektury zorientowanej na użytkownika występuje największy zysk względem architektury zorientowanej na sieć w przypadku kwantyla rzędu 10 reprezentującego użytkowników o najgorszych warunkach propagacyjnych. Niestety w przypadku mediany i kwantyla rzędu 90 reprezentującego użytkowników o najlepszych warunkach propagacyjnych można zaobserwować znaczące pogorszenia odpowiednio aż o 44\% i 47\%. Z drugiej strony, zaproponowany algorytm poprzez odpowiednie formowanie klastrów obsługujących, pozwala na zwiększenie zarówno kwantyla rzędu 10 jak i mediany z rozkładu przepływności użytkowników odpowiednio o 51\% i 16\% względem architektury zorientowanej na sieć. Zysk osiągnięty jest kosztem minimalnego pogorszenia kwantyla rzędu 90, tj., o 4\%.
\section{wnioski} \label{sec:conclusions}

Jednym z najważniejszych wyzwań związanych z implementacja sieci Open RAN o architekturze zorientowanej na użytkownika jest odpowiednie formowanie klastrów obsługujących. Problem ten może być efektywnie rozwiązany za pomocą zaproponowanego algorytmu wykorzystującego współpracę pomiędzy xApp'em i rApp'em. Wyniki symulacji pokazały, że zaproponowane rozwiązanie pozwala na zwiększenie mediany z rozkładu przepływności użytkowników odpowiednio o 16\% i 52\% względem architektury zorientowanej na sieć i kanonicznej implementacji architektury zorientowanej na użytkownika.

\bibliographystyle{krit}
\bibliography{references}
\end{multicols}
\end{document}